%% file: paper.tex
\documentclass[runningheads]{llncs}
\usepackage[T1]{fontenc}
\usepackage{graphicx}
\usepackage{booktabs}
\usepackage[misc]{ifsym}

\usepackage{enumitem}
\usepackage{tabularx}
\usepackage{array}
\usepackage{ragged2e}
\usepackage{tcolorbox}
\usepackage{subcaption}
\usepackage{xparse}
\usepackage{xcolor}
\usepackage{colortbl}
\usepackage{threeparttable}
\usepackage{float}
\usepackage[section]{placeins}

\input{commands}
\usepackage{multirow}
\begin{document}

\title{Same Outcomes, Different Journeys: A Trace-Level Framework for Comparing Human and GUI-Agent Behavior in Production Search Systems}

\titlerunning{Same Outcomes, Different Journeys}

\author{Maria Movin\inst{1,3} \and Claudia Hauff\inst{2} \and Aron Henriksson\inst{3} \and Panagiotis Papapetrou\inst{3}}

\authorrunning{M. Movin et al.}

\institute{Spotify, Sweden \and Spotify, Netherlands \and Stockholm University, Sweden}

\maketitle

\begin{abstract}
\input{sections/0_abstract}
\end{abstract}

\input{sections/1_intro_ecml_pkdd}
\input{sections/2_related_work_acl}
\input{sections/3_method}
\input{sections/5_results}
\input{sections/7_practical_lessons}
\input{sections/8_limitations}
\input{sections/9_conclusion}

\bibliographystyle{splncs04}
\bibliography{references}
\end{document}

%% file: commands.tex
\newcommand{\appagent}[0]{\texttt{AppAgent}}
\newcommand{\noncompliant}[1]{$\dagger$#1}
\newcommand{\gptknow}[1]{$\ddagger$#1}

\definecolor{agentcolor}{RGB}{102,194,165}
\definecolor{usercolor}{RGB}{231,138,195}

%% file: sections/0_abstract.tex
LLM-driven GUI agents are increasingly used in production systems to automate workflows and simulate users for evaluation and optimization. Yet most GUI-agent evaluations emphasize task success and provide limited evidence on whether agents interact in human-like ways. We present a trace-level evaluation framework that compares human and agent behavior across (i) task outcome and effort, (ii) query formulation, and (iii) navigation across interface states. We instantiate the framework in a controlled study in a production audio-streaming search application, where 39 participants and a state-of-the-art GUI agent perform ten multi-hop search tasks. The agent achieves task success comparable to participants and generates broadly aligned queries, but follows systematically different navigation strategies: participants exhibit content-centric, exploratory behavior, while the agent is more search-centric and low-branching. These results show that outcome and query alignment do not imply behavioral alignment, motivating trace-level diagnostics when deploying GUI agents as proxies for users in production search systems.
\keywords{GUI Agents \and Multi-hop Search \and Human--Agent Behavior}

%% file: sections/1_intro_ecml_pkdd.tex
\section{Introduction}\label{sec:intro}

Recent progress in multimodal GUI agents has enabled LLMs to interact directly with applications by perceiving visual interface states and generating action sequences~\cite{cheng2024seeclick,nguyen2024gui,appagent}. These agents can now perform complex workflows in real environments, including information-seeking tasks that involve issuing queries, navigating results, and integrating information across interface transitions.

Existing GUI-agent benchmarks primarily evaluate capabilities such as element grounding and end-to-end task completion~\cite{cheng2024seeclick,deng2023mind2web,koh2024visualwebarena,xie2024osworldbenchmarkingmultimodalagents}. While essential, these benchmarks provide limited insight into the interaction strategies agents employ while navigating systems. This distinction becomes particularly important when GUI agents are used as \emph{proxies} for \emph{human users} to evaluate or optimize information-seeking systems. Such proxy use is increasingly explored in practice, for example through agent-simulated A/B-style comparisons of interface variants \cite{wang2025agenta} and piloting usability studies with simulated users~\cite{lu2025uxagent}. If agents reach the same outcomes as humans but follow systematically different interaction strategies, systems evaluated or optimized on agent traces may learn behaviors that users do not exhibit. For example, consider a music‑app task where the goal is to ``find an artist similar to Artist X who released an album after 2020.'' A state-of-the-art GUI agent may handle this by repeatedly reformulating a search query, scanning results, and selecting the first matching entity. A human user, however, might navigate to Artist X’s profile, browse related artists, inspect discographies, follow links to album pages, and backtrack across views before choosing a candidate. Although both may succeed, their navigation strategies diverge in ways that remain invisible under outcome‑based evaluation alone.

To address this gap, we propose a trace-level evaluation framework for comparing human and agent behavior in GUI-based information-seeking tasks across three complementary dimensions: (i) task outcomes and effort, (ii) query formulation patterns, and (iii) navigation behavior. We demonstrate the framework in a controlled human--agent study in a production audio-streaming application, where 39 participants and a state-of-the-art GUI agent~\cite{appagent} perform ten multi-hop information-seeking tasks requiring query reformulation, entity discovery, and backtracking across interface states~\cite{mavi2024multi}. Unlike most evaluations of GUI agents, our study uses real user accounts and production interaction logs, providing ecological validity rarely available in GUI agent studies.

Our analysis shows that the agent achieves task success rates comparable to human participants while requiring fewer actions but substantially more time. Agent-generated queries broadly align with participant queries, suggesting similar expressions of information need. However, navigation behavior diverges: the agent follows search-centric, low-branching trajectories, whereas participants engage in more content-centric and exploratory navigation. These findings show that alignment in outcomes and language does not imply behavioral alignment, and that trace-level analysis can reveal differences that remain hidden under success-based evaluation alone. Our \textbf{contributions} are as follows:
\begin{itemize}
\item \textbf{Novelty.} We present a structured trace-level evaluation framework for assessing behavioral alignment between GUI agents and human users in production search systems, enabling analysis beyond outcome-based evaluation.
\item \textbf{Real-world validation.} We perform a controlled human–agent comparison in a live, personalized production search application, demonstrating the framework under realistic deployment conditions. This setting allows us to capture authentic user behavior and model interactions that cannot be replicated in synthetic sandbox environments.
\item \textbf{Practical insight.} We provide evidence that outcome and query alignment can mask navigation-level divergence, motivating behavior-aware validation when agent traces inform optimization in production search systems. Our findings highlight specific behavioral mismatches that may influence system tuning, recommendations, and interface design.
\end{itemize}

%% file: sections/2_related_work_acl.tex
\section{Related Work}
\subsection{GUI Agents}
LLMs combined with multimodal perception have enabled GUI agents that operate directly on visual interfaces, perceiving screenshots and issuing actions to complete tasks~\cite{cheng2024seeclick,nguyen2024gui,appagent}. Earlier approaches relied on structured interface representations, whereas contemporary \emph{visual} agents reason jointly over pixels, layout, and text, and plan sequences of clicks, swipes, and text inputs with the help of application-specific harnesses and offline pretraining~\cite{cheng2024seeclick,appagent}.

Recent evaluations of GUI agents rely on benchmarks such as Mind2Web~\cite{deng2023mind2web}, VisualWebArena~\cite{koh2024visualwebarena}, and OSWorld~\cite{xie2024osworldbenchmarkingmultimodalagents}, which simulate complex tasks in controlled environments. Although these benchmarks log interaction traces, evaluation largely centers on task completion, offering limited insight into how agents interact with interfaces or how their strategies compare to those of human users. Because these environments rely on controlled setups—such as offline caches~\cite{deng2023mind2web}, self-hosted mockups~\cite{koh2024visualwebarena}, or fixed initial states~\cite{xie2024osworldbenchmarkingmultimodalagents}—it remains unclear how agents behave in live systems with dynamic content and personalized interfaces.

Beyond benchmarks, recent work explores using LLM-driven agents as scalable proxies for interface evaluation, including agent-based usability testing for web design~\cite{lu2025uxagent} and agent-simulated A/B-style comparisons of interface variants~\cite{wang2025agenta}. These approaches motivate the need to understand not only whether agents \emph{succeed}, but whether their \emph{interaction behavior} reflects that of human users when agent traces are used to inform design or system decisions. Complementary work compares human and agent workflows for multi-step computer tasks~\cite{wang2025workflow}. Although agents sometimes resemble human workflows, they rely heavily on programmatic tools and lag behind humans in success and work quality. However, this abstraction offers limited insight into query formulation and navigation across interface states in deployed information-seeking systems.

We address these gaps by introducing a trace-level framework for comparing human and agent behavior in GUI-based information-seeking tasks, enabling analysis of how their interaction strategies differ when performing the same tasks within a production system.

\subsection{User Simulation and Search}

User simulation has long been used in information retrieval to study how user
behavior affects system evaluation and design~\cite{balog2024fntirusersim}.
Traditional simulators rely on explicit behavioral assumptions about querying,
clicking, and stopping~\cite{keskustalo2008evaluating,maxwell2016simulating,thomas2014modeling},
often calibrated using user logs or controlled studies. More recently, LLM-based
approaches generate search traces from natural-language task descriptions,
providing a flexible alternative to hand-crafted behavioral models
~\cite{zhang2024usimagent,zhang2025humanlikethinking}.

With the emergence of LLM-driven GUI agents that operate directly on application interfaces, user simulation can shift from generating abstract query--click sequences to producing end-to-end interaction traces that include navigation across heterogeneous interface states. This shift raises new questions about behavioral fidelity beyond outcome-based evaluation. We investigate this by comparing human and agent traces in a production search application, focusing on query formulation and state-level navigation.

%% file: sections/3_method.tex
\begin{figure}[t]
    \centering
    \includegraphics[width=\linewidth]{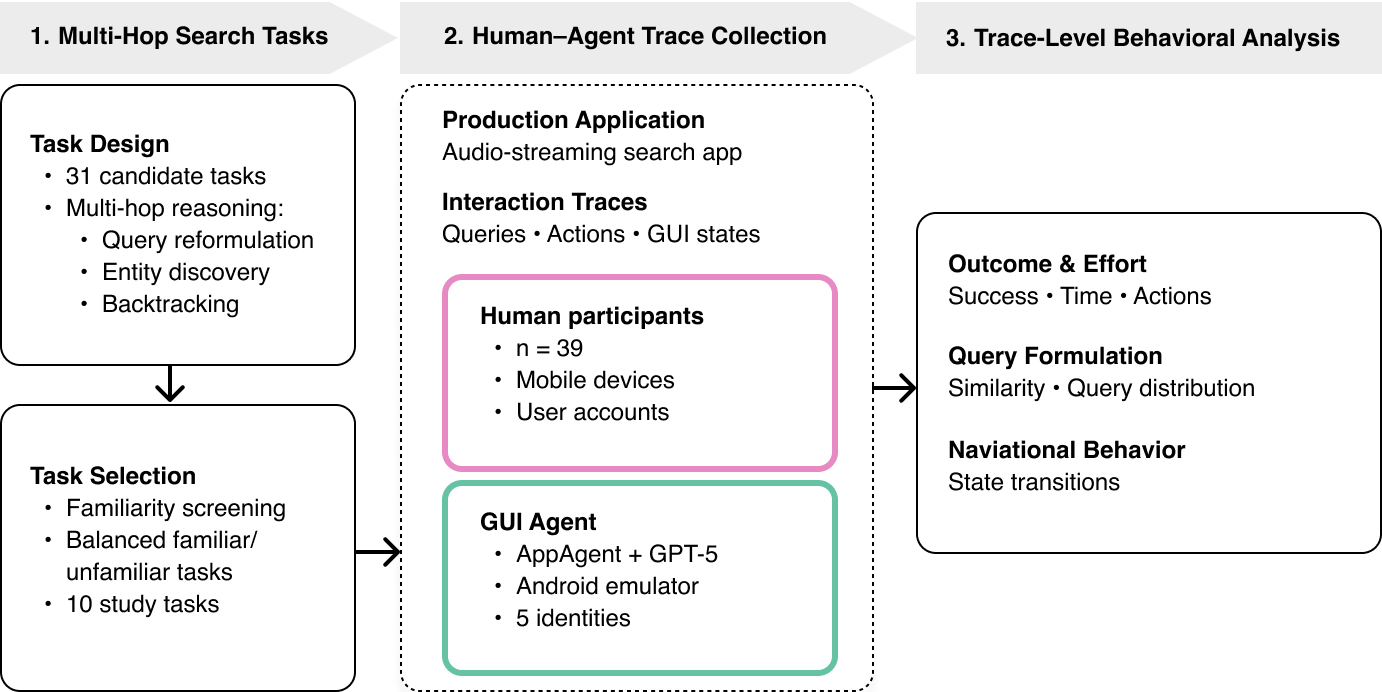}
    \caption{Overview of the trace-level evaluation framework: task design, human–agent trace collection in a production search application, and behavioral comparison across outcomes, query formulation, and navigation metrics.}
    \label{fig:evaluation_framework}
\end{figure}
\section{Trace-Level Evaluation Framework}

We study human–agent alignment using a trace-level evaluation framework for GUI-based information-seeking tasks. As illustrated in Figure~\ref{fig:evaluation_framework}, the framework consists of three components: (a) designing multi-hop information-seeking tasks that require combining information across search results and entity pages~\cite{mavi2024multi}, (b) collecting aligned human and agent interaction traces during task execution within the same application environment, and (c) comparing cohorts along three complementary behavioral dimensions: (i) task outcomes and effort, (ii) query formulation patterns, and (iii) navigation behavior across interface states.

We instantiated this framework in the Spotify application, a production audio-streaming application that provides a search interface over music and podcast entities.

Since the application was familiar to participants, no additional training was required, allowing us to observe behavior in a realistic information-seeking environment. To account for personalization effects, agent experiments were executed under multiple user identities.

\input{tables/task_descriptions_only_tasks}

\subsection{Search Tasks and Familiarity Screening}
We created an initial set of 31 multi-hop search tasks centered around music and podcast entities. We used the following criteria during this process: (i) each task requires at least one intermediate reasoning step (a query reformulation, entity bridging, or in-application backtracking), (ii) each task can be solved entirely within the application interface, and (iii) each task has an objectively verifiable success criterion (i.e., playing a specific track or any track from a defined set). Each of our tasks followed one of two reasoning patterns: \textit{linear chains}, where the solution requires a sequence of \emph{dependent} steps, or \textit{entity bridging}, where information across different entities has to be connected.

Using convenience sampling from the authors’ immediate surroundings, including academic and industry contacts, we recruited 47 participants. To control for prior knowledge in the search tasks, we conducted a familiarity screening before the study. Each participant completed a short online questionnaire assessing their familiarity with each candidate task\footnote{
We slightly reformulated tasks into screening questions; e.g., the task T3 ``\emph{Play the first song of an album released in 2023 by one of the singers currently twice in the top-10 of the UK singles charts}'' was presented as ``\emph{What is the 2023 album of the singer currently twice in the top-10 of the UK singles charts?}''
}. The response options were \emph{I am familiar with some entities}, \emph{I know the answer to the question}, and \emph{I don’t know anything}. Tasks marked \emph{I know the answer to the question} were excluded to avoid solutions derived solely from prior knowledge. The remaining responses were used to categorize tasks as \emph{familiar} or \emph{unfamiliar}. \emph{Familiar} tasks involve recognizable entities but still require search within the application, whereas \emph{unfamiliar} tasks require entirely in-application exploration.

We ranked the 31 candidate tasks by the balance of familiarity responses and selected the smallest subset that enabled assigning each participant exactly two \emph{familiar} and two \emph{unfamiliar} tasks while maintaining a balanced distribution. Using the smallest feasible subset also increased the number of observations per task, resulting in a final set of ten tasks (Table~\ref{tab:tasks-grouped}).

\subsection{Participants' Search Task Execution}
A week after the initial questionnaire, we invited participants to the second part of the study. They used their own mobile devices to complete the search tasks while logged into their personal user accounts, allowing personalized search functionality to remain active. Participants were guided through the study via a personalized web form. After providing consent for the collection and analysis of session logs, they completed a questionnaire on demographics, application usage, and search experience. This was followed by a short training task to familiarize participants with the study procedure. Each participant was then assigned four tasks in a random order (two \emph{familiar}, two \emph{unfamiliar}).

Before starting each task, participants rated the anticipated task difficulty on a four-point scale (Table~\ref{tab:tasks-execution}, column \textbf{TD}). They were instructed to use only the audio-streaming application (no external sources) and to treat 10 minutes as a soft per-task limit, skipping the task if the time elapsed without success. For each task, we recorded task-level start and end times; client-side instrumentation\footnote{Note that the framework does not require custom instrumentation. Screen recordings of the device are sufficient to reconstruct the traces (queries and navigational states) and determine outcomes from the visible UI.}
captured event-level logs (actions, elements/URIs, timestamps).
We used the final played track as the outcome: success if it satisfied the task,
otherwise failure.

In total, 39 participants completed the second step of the study, yielding 150 logged participant-task combinations (6 tasks were not successfully logged, or not executed by the respective participant). Most participants were long-term users of the audio-streaming application (69\% $>10$ years, 28\% 5–10 years, 3\% less than a year), with varied search usage frequency in the application (35\% 1–3 times per week, 25\% less than once a week, 25\% about once a day, 15\% 2–5 times a day). Participants were predominantly based in Sweden (33/39, 85\%), with smaller groups from the UK (4/39, 10\%) and other countries (2/39, 5\%). The interface languages were English (23/39, 59\%) and Swedish (16/39, 41\%). To capture prior experience, we estimated each participant’s total number of queries based on reported years of application usage and search frequency. Based on this estimate, we classified 16 participants (41\%) as \emph{expert users} and 23 (59\%) as \emph{regular users}. These cohorts were used in subsequent analyses.

\subsection{GUI Agent Search Task Execution}

To instantiate the framework, we executed the same search tasks using a GUI agent interacting with the production audio-streaming application. Our goal was to assess whether a strong GUI agent exhibited human-like interaction behavior under the same task conditions as participants, rather than to compare different agent architectures.

After experimenting with several GUI-agent frameworks and LLMs, we selected \appagent{}~\cite{appagent} with GPT-5~\cite{openai2025gpt5}, which provided the most reliable navigation and instruction following in the application. The agent used the \texttt{gpt-5-2025-\\08-07} model with reasoning effort set to \texttt{medium}, temperature 0.0, and a maximum completion length of 7,000 tokens. Pilot experiments showed that the medium reasoning setting achieved task success and instruction adherence comparable to higher reasoning configurations while substantially reducing latency. All runs were executed in an Android emulator using the same application interface as participants.

During pilot experiments, we observed that models often attempted to answer tasks from prior knowledge rather than deriving answers through interaction with the application. Notably, GPT-5 could answer 7 of the 10 tasks without interacting with the interface (Table~\ref{tab:tasks-execution}, marked with \gptknow{} in column \textbf{ID}). To enforce in-application reasoning, we introduced prompt constraints requiring the agent to obtain answers through exploration within the interface (Fig.~\ref{fig:prompt}). Although these constraints substantially reduced shortcut solutions, occasional violations still occurred. Agent runs were therefore audited for compliance, and non-compliant runs were filtered from the main analysis (described below). Smaller models or lower-reasoning configurations violated this constraint more frequently in pilot experiments, further motivating the use of a high-capacity model.

\subsubsection{Offline Exploration Phase.}

Following AppAgent setup procedures \cite{appagent}, we performed an offline exploration phase using five unused multi-hop search tasks. The agent generated a structured representation of UI elements and their functions, which was reused during subsequent runs.

\subsubsection{Runs and Compliance Filtering.}

We executed 50 agent runs by performing each of the ten tasks with five synthetic test identities. Because the application provides a personalized experience, the test identities were configured to cover the participant pool’s locales and interface settings (3 Sweden, 1 US, 1 UK; 3 Swedish, 2 English). Each run was capped at 50 actions; runs exceeding this limit were considered unsuccessful.

Agent logs were manually audited for compliance with the in-application reasoning constraint (Fig.~\ref{fig:prompt}), requiring answers to be derived from interface interactions rather than prior knowledge or external tools. We flagged 11 of the 50 runs as non-compliant. Two patterns emerged: three isolated violations (T4, T6, T7) and systematic shortcut behavior on T9 and T10 (4/5 runs each)\footnote{For example, T4 requires discovering the name of a New York Times podcast through search, but the initial query “popcast” directly reveals the answer.}. We defined two analysis sets: the \textbf{Compliance Set (CS)}, excluding flagged runs, and the \textbf{Task-Excluded (TE)} set, which removes T9 and T10 because only one compliant run remains per task.

\input{prompt_instructions_2}

\subsection{Analysis Framework and Metrics}

The trace-level evaluation framework compares human users and GUI agents across three dimensions: (i) task outcomes and effort, (ii) query formulation patterns, and (iii) navigation behavior. All analyses use the filtered CS and TE sets defined above. Unless otherwise stated, metrics are computed per run and aggregated across tasks. Participant-only subgroup analyses compare expert versus regular users and \emph{familiar} versus \emph{unfamiliar} tasks.

\subsubsection{Task Outcomes and Effort.}
Outcome and effort comparisons use the CS set. Task success is determined from the logged outcome per run. Effort is measured by the number of actions and total task time. The same run-level measures are used for participant subgroup comparisons, and all agent runs, including filtered ones, are inspected to characterize shortcut behavior.

\subsubsection{Query Formulation Patterns.}

\textit{First-query similarity.}
Using the TE set, we extract the \emph{first query} for each run and apply light normalization (lowercasing and whitespace removal). For each task, we compute pairwise similarity between agent--participant, participant--participant, and agent--agent first-query pairs using Python's \texttt{SequenceMatcher}, and macro-average these task-level means across tasks. As a robustness check, we replicate the same comparison in a TF--IDF vector space using cosine similarity. \texttt{SequenceMatcher} captures character-level similarity between query strings, whereas TF--IDF cosine measures overlap in informative terms. We use lexical metrics to compare query formulations rather than semantic equivalence. This choice is appropriate for short, entity-oriented queries typical in entity search. We use the same first-query framework for participant subgroup comparisons.

\paragraph{Full-query distributional analysis.}
Using the TE set, we represent the full set of user and agent queries in a shared TF--IDF space using unigrams and bigrams, with unique user queries weighted by frequency. We then compute \emph{typicality} relative to the frequency-weighted user centroid, \emph{coverage} as the share of user query mass whose nearest agent query exceeds a cosine-similarity threshold $\tau$, and \emph{efficiency} as cumulative coverage as unique agent queries are added. Coverage is evaluated over multiple values of $\tau$, and the efficiency analysis is illustrated at $\tau=0.6$. We compare the resulting efficiency curve against two baselines: a size-matched random subset of unique participant queries and a top-$N$ frequency oracle consisting of the $N$ most frequent participant queries.

\subsubsection{Navigation Behavior.}

Navigation destinations are mapped to a reduced set of semantic GUI states
(e.g., search, artist, playlist, now-playing). Directed
transition graphs are then built for users and agents, where nodes represent GUI states and
edges represent observed transitions, weighted by
frequency.

To quantify shared navigation structure, we compare the cohorts' top-$k$ transitions using Jaccard overlap:
\[
J=|E^{(\mathrm{agent})}_{k}\cap E^{(\mathrm{user})}_{k}|/|E^{(\mathrm{agent})}_{k}\cup E^{(\mathrm{user})}_{k}| ,
\]
where $E_k$ denotes the set of the $k$ most frequent directed transitions in a cohort-specific graph. We examine multiple values of $k$: smaller values highlight the shared high-frequency backbone, whereas larger values incorporate lower-frequency, more task-specific transitions. In the reported results, we highlight $k=10$ and $k=20$. We report both \emph{pooled} (\emph{micro}) overlap, in which all runs are combined before graph construction, and \emph{task-equal} (\emph{macro}) overlap, in which overlap is computed per task and then averaged across tasks. Navigation comparisons use both the CS and TE sets: pooled overlap is reported on CS, whereas task-equal overlap is reported on TE.

\input{tables/success_per_task}

\subsubsection{Statistical Analysis.}
We report proportions with 95\% confidence intervals for success measures.  Differences in success proportions between cohorts are tested using Pearson’s $\chi^2$ tests on pooled success counts. For continuous or count-based effort measures, we use two-sided Mann--Whitney U tests (MWU) for cohort/subgroup comparisons; for the family of effort comparisons reported together, we check robustness under Holm--Bonferroni correction.  For query-similarity analyses, we report macro-averaged task-level means with bootstrap 95\% confidence intervals; subgroup comparisons use MWU tests, and uncertainty around the size-matched random baseline in the full-query analysis is estimated with bootstrap confidence intervals.

%% file: tables/task_descriptions_only_tasks.tex
\newcolumntype{Y}{>{\RaggedRight\arraybackslash}X}

\definecolor{agentcolor}{RGB}{102,194,165}
\definecolor{usercolor}{RGB}{231,138,195}

\newcommand{\TaskRow}[2]{#1 & #2}

\begin{table}[htb!]
  \centering
  \setlength{\tabcolsep}{3pt}
    \renewcommand{\arraystretch}{0.98}

  \begin{threeparttable}
\caption{Multi-hop search tasks by reasoning pattern (T1--T10)
}
  \label{tab:tasks-grouped}

\begin{tabularx}{\linewidth}{@{}l Y@{}}
    \toprule
    \textbf{ID} & \textbf{Task Instruction}
    \\
    \midrule

    \multicolumn{2}{@{}c}{\textit{Linear}} \\
    \midrule

    \TaskRow{T1}{Find the artist who was guest at the podcast ``Sommar i P1'' 30~June~2023, and play one of his tracks}\\
    \TaskRow{T2}{Find the artist who has covered the song with the lyrics `Jag vill känna att jag lever', originally performed by Helen Sjöholm, and play a track from their latest album}\\
    \TaskRow{T3}{Play the first song of an album released 2023 of one of the singers who is currently twice in the top-10 of the UK singles charts}\\
    \TaskRow{T4}{Play the most streamed song of the artist interviewed in the \emph{New York Times} podcast about pop music from 22~August}\\
    \TaskRow{T5}{Play the first track of the newest album of the composer writing the soundtrack for a Netflix series where chess features heavily}\\
    \TaskRow{T6}{Brandy and Monica recorded a duet, one of today's most globally popular artists has released a track with the same title. Play one of the latest releases by this artist, created together with her brother.}\\
    \TaskRow{T7}{Since 2020 the video game \emph{Minecraft} has a new main composer, earlier known for her soundtrack work on \emph{Celeste}. She also produces electronic music under another alias. Play her most streamed track released under her alias}\\

    \midrule
    \multicolumn{2}{@{}c}{\textit{Entity bridging}} \\
    \midrule
    \TaskRow{T8}{Play a song from the actress that sang songs with both Robbie Williams and Luke Evans}\\
    \TaskRow{T9}{Play the song covered by Maria Callas that also appears in the soundtrack of the original \emph{Fantasia} movie}\\
    \TaskRow{T10}{The composer of the hit video game \emph{God of War} (2018) has also created a solo rock concept album. From that album play the song featuring the artist singing ``Chop Suey!''}\\

    \bottomrule
  \end{tabularx}

  \end{threeparttable}

\end{table}

%% file: prompt_instructions_2.tex
\begin{figure}[t]
\centering
\begin{tcolorbox}[
  width=\columnwidth,
  colback=gray!10, colframe=gray!50,
  boxrule=0.4pt, arc=2pt,
  left=8pt, right=8pt, top=6pt, bottom=6pt
]
\scriptsize
\textbf{YOU HAVE NO EXTERNAL KNOWLEDGE.} You are completely ignorant about:
\begin{itemize}[leftmargin=1.4em,itemsep=0.15em,topsep=0.2em]
  \item Song titles, artists, albums, or any music information
  \item Movie names, actors, directors, or entertainment facts
  \item Historical events, dates, or general knowledge
  \item Any information not explicitly visible in the current screenshot
\end{itemize}

\textbf{YOU MUST DISCOVER ALL INFORMATION THROUGH THE APP:}
\begin{itemize}[leftmargin=1.4em,itemsep=0.15em,topsep=0.2em]
  \item To find who wrote a song, first search for that song in the app and examine the credits/details shown
  \item To find the most popular/streamed item, examine numbers, rankings, or popularity indicators visible in the app
  \item To find connections between items, look up each item separately in the app interface
  \item Never assume facts that are not displayed on screen
\end{itemize}

\textbf{STAY WITHIN THE AUDIO STREAMING APP:}
\begin{itemize}[leftmargin=1.4em,itemsep=0.15em,topsep=0.2em]
  \item Operate within the audio mobile app only
  \item If a link redirects to a browser or external app, immediately return to the app
  \item Complete all tasks within the audio app interface; do not use external applications
\end{itemize}

If information is not currently visible, your \textbf{only} option is to search for
and navigate to that information within the app.
\end{tcolorbox}

\caption{Prompt instructions added to the \appagent{} system prompt to encourage in-application reasoning rather than answering from prior knowledge.}
\label{fig:prompt}
\end{figure}

%% file: tables/success_per_task.tex
\newcolumntype{Y}{>{\RaggedRight\arraybackslash}X}

\renewcommand{\TaskRow}[8]{#1 & #2 & #3 & #4 & #5 & #6 & #7 & #8 \\}

\newcommand{\QThree}[3]{
  \begingroup
    \renewcommand{\arraystretch}{0.95}
    \begin{tabular}[t]{@{}p{\linewidth}@{}}
      \RaggedRight #1\\#2\\#3
    \end{tabular}
  \endgroup
}

\renewcommand{\QThree}[3]{
  \begingroup
    \renewcommand{\arraystretch}{0.95}

    \pretolerance=10000
    \tolerance=2000
    \emergencystretch=2em
    \linepenalty=1000
    \hyphenpenalty=10000
    \exhyphenpenalty=10000
    \begin{minipage}[t]{\linewidth}
      \RaggedRight
      #1\par
      #2\par
      #3
    \end{minipage}
  \endgroup
}

\newcommand{\UQ}[3]{\QThree{#1}{#2}{#3}}

\newcommand{\QThreeAgent}[3]{
  \begingroup
    \renewcommand{\arraystretch}{0.95}
    \parbox[t]{\hsize}{
      \RaggedRight
      \tolerance=1500
      \emergencystretch=0.5em
      \linepenalty=50
      #1\par #2\par #3
    }
  \endgroup
}

\newcommand{\QThreeUser}[3]{
  \begingroup
    \renewcommand{\arraystretch}{0.95}
    \parbox[t]{\hsize}{
      \RaggedRight
      \tolerance=9000
      \emergencystretch=4em
      \linepenalty=100000
      \hyphenpenalty=10000
      \exhyphenpenalty=10000
      #1\par #2\par #3
    }
  \endgroup
}

\newcommand{\AQ}[3]{\QThreeAgent{#1}{#2}{#3}}

\begin{table}[H]
\centering
\setlength{\tabcolsep}{4pt}
\renewcommand{\arraystretch}{1.05}
\begin{threeparttable}
\caption{Multi-hop tasks and execution statistics for agent runs and participants.}
\label{tab:tasks-execution}

\begin{tabularx}{\linewidth}{@{}
      p{0.045\linewidth}
  >{\columncolor{agentcolor!20}\centering\arraybackslash}p{0.045\linewidth}
  >{\columncolor{agentcolor!20}\hsize=1.05\hsize}Y
  >{\columncolor{usercolor!20}\centering\arraybackslash}p{0.055\linewidth}
  >{\columncolor{usercolor!20}\centering\arraybackslash}p{0.035\linewidth}
  >{\columncolor{usercolor!20}\centering\arraybackslash}p{0.0255\linewidth}
  >{\columncolor{usercolor!20}\centering\arraybackslash}p{0.0455\linewidth}
  >{\columncolor{usercolor!20}\hsize=0.95\hsize}Y
@{}}
\toprule
\multicolumn{1}{@{}l}{} &
\multicolumn{2}{c}{\textbf{Agent}} &
\multicolumn{5}{c@{}}{\textbf{Participant}}\\[-2pt]
\cmidrule(l){2-3}\cmidrule(l){4-8}

\textbf{ID} &
\textbf{\rotatebox[origin=c]{50}{\shortstack{S/$n_c$}}} &

\textbf{Queries} &
\textbf{\rotatebox[origin=c]{50}{\shortstack{S/n}}} &
\textbf{\rotatebox[origin=c]{50}{Fam}} &
\textbf{\rotatebox[origin=c]{50}{TD}} &
\textbf{\rotatebox[origin=c]{50}{\shortstack{t (s)}}} &
\textbf{Queries} \\
\midrule
\TaskRow{T1}{2/5}{\AQ{sommar i p1}{sommar i p1s}{sommar i p1 30 june 2023}}{11/18}{11}{2.4}{$230\pm182$}{\UQ{sommar i p1}{sommar i p1 2023}{guest in sommar i p1}}
\TaskRow{\gptknow{T2}}{3/5}{\AQ{jag vill känna att jag}{jag vill känna att j}{jag vill känna}}{12/18}{12}{2.4}{$108\pm86$} {\UQ{jag vill känna att jag lever}{gabriellas sång}{cover på jag vill känna att jag lever}}
\TaskRow{T3}{4/5}{\AQ{top 50 uk}{uk singles chart}{uk top 50}}{10/14}{6}{2.9}{$171\pm180$}{\UQ{top 10 uk}{uk top 10 singles}{uk singles}}
\TaskRow{T4}{2/4}{\AQ{new york times pop music podcast}{new york times pop music}{\noncompliant{popcast}}}{8/15}{8}{3.0}{$218\pm137$}{\UQ{new york times}{new york times podcast}{which artist was interviewed in new york times podcast}}
\TaskRow{\gptknow{T5}}{3/5}{\AQ{chess netflix}{netflix chess soundtrack}{netflix chess series}}{7/12}{6}{2.6}{$110\pm103$}{\UQ{queens gambit}{netflix}{netflix series chws}}
\TaskRow{\gptknow{T6}}{0/4}{\AQ{brandy and monica}{brandy monica duet}{\noncompliant{the boy is mine}}}{1/13}{6}{2.7}{$284\pm139$} {\UQ{brandy and monica}{brandy and monica duet}{the boy is mine}}
\TaskRow{\gptknow{T7}}{2/4}{\AQ{minecraft composer 2020 celeste}{celeste soundtrack}{\noncompliant{lena raine}}}{9/15}{7}{3.5}{$228\pm173$} {\UQ{minecraft}{minecraft composer}{celeste soundtrack}}
\TaskRow{\gptknow{T8}}{4/5}{\AQ{robbie williams}{}{}}{9/17}{9}{3.1}{$199\pm225$}{\UQ{robbie williams luke evans}{robbie williams}{robbie williams, luke evans and}}
\TaskRow{\gptknow{T9}}{1/1}{\AQ{fantasia soundtrack}{}{}}{7/15}{6}{2.6}{$251\pm232$}{\UQ{fantasia}{maria callas}{maria callas fantasia}}
\TaskRow{\gptknow{T10}}{1/1}{\AQ{god of war 2018}{god of war}{god of war 2018 co}}{6/13}{4}{2.5}{$309\pm151$}{\UQ{god of war}{god of}{chop suey}}

\bottomrule
\end{tabularx}

\begin{tablenotes}[para,flushleft]
\item \textbf{ID}: agent had internal knowledge on tasks marked with \gptknow{}.
\item \textbf{S}: number of successful completions.
\item \textbf{$n_c$}: number of compliant attempts.

\item \textbf{n}: number of participants.
\item \textbf{Fam}: number of familiar participants.
\item \textbf{TD}: mean task difficulty rating (1--4).
\item \textbf{t (s)}: seconds (mean$\pm$SD).
\item \textbf{Queries}: three representative queries. Non-compliant queries are marked with \noncompliant{} and are not included in CS/TE analyses.
\end{tablenotes}

\end{threeparttable}
\end{table}

%% file: sections/5_results.tex
\section{Results}

\subsection{Task Outcomes and Effort}

Table~\ref{tab:effort-metrics} summarizes success and effort. Success rates were comparable between agent runs and participants (56.4\% vs.\ 53.3\%). However, agent runs were much slower despite fewer actions  (mean $\pm$ SD: $1076 \pm 760$ vs.\ $209 \pm 175$ seconds; actions median 11 [9, 19] vs.\ 18 [12, 30]; both MWU tests $p<.001$). Agent time includes model inference and emulator latency and was not optimized, so slower runtimes mainly reflect system overhead rather than interaction inefficiency. Performance varied considerably by task (Table~\ref{tab:tasks-execution}); among T1--T8, no task was solved by all agent runs or participants. T6 was the most difficult\footnote{The application content changed between task design and study execution, making the correct track harder to find even though the task itself remained valid.}, with no successful agent runs and only one successful participant attempt.

We next examined differences within the participant cohort. Experts had a higher success rate than regular participants (62.9\% vs.\ 46.6\%), although the difference was not statistically significant. Experts completed tasks with lower effort (time $150 \pm 107$ vs.\ $251 \pm 200$ s, MWU $p=.001$; actions median 18 [12, 25] vs.\ 21 [15, 38], MWU $p=.05$). Task familiarity showed a similar pattern. Success rates were numerically higher on familiar tasks (60.0\% vs.\ 46.7\%) but not statistically significant, whereas participants completed familiar tasks faster and with fewer actions (time $184 \pm 172$ vs.\ $237 \pm 177$ s, MWU $p<.01$; actions median 19 [13, 32] vs.\ 21 [14, 38], MWU $p<.01$). This suggested that familiarity primarily reduced interaction effort rather than determining task success. Even when entities were recognizable, participants still relied on the application to locate and verify the correct content. All significance conclusions remained unchanged after Holm–Bonferroni correction across the subgroup effort comparisons.

Relative to these participant patterns, agent success (56.4\%) fell between the expert–regular and familiar–unfamiliar participant ranges. In terms of actions, agent runs were closer to the lower-effort patterns observed for expert users and familiar tasks. Finally, we examined the effect of filtering non-compliant runs. Including all 50 agent runs increased success from 56.4\% to 62.0\%. The excluded runs required substantially fewer actions (median 7 [4, 7] vs. 11 [9, 19]) and completed faster (mean 463 s vs.\ 1076 s), indicating shortcut solutions that bypassed in-application reasoning. We therefore reported results on the CS and TE analysis sets to isolate comparable GUI interaction behavior.

\input{tables/effort}

\subsection{Query Formulation Patterns}
\paragraph{First-query similarity.}
For the same task, the agent’s initial query was about as similar to a participant’s query as two participants’ queries were to each other. Macro-averaged across tasks, agent–participant \texttt{SequenceMatcher} similarity was 0.58 (95\% CI [0.51, 0.64]), comparable to participant–participant similarity of 0.57 (95\% CI [0.45, 0.66]).
Agent–agent similarity was higher at 0.75 (95\% CI [0.62, 0.88]), indicating more consistent query formulations across agent runs.
A TF–IDF cosine replication yielded the same ordering (agent–agent $>$ agent–participant $\approx$ participant–participant), indicating metric robustness.

For illustration, T1 (Swedish podcast) had the highest \texttt{SequenceMatcher} similarity (mean 0.72), with near-identical queries beginning with ``\emph{Sommar i P1}'', while T5 (Netflix/chess) had the lowest (0.28): agent runs used generic terms (e.g., ``\emph{chess netflix}'') while some participants relied on prior knowledge (e.g., ``\emph{queen's gambit}''), which the agent was instructed to avoid (cf. Table~\ref{tab:tasks-execution}).

Experts’ initial queries aligned more with agent runs than those of regular participants (Expert–Agent query similarity $0.61$ vs.\ Regular–Agent $0.55$; MWU $p=0.442$), a non-significant trend consistent with higher expert–expert similarity ($0.65$ vs.\ regular–regular $0.53$). Task familiarity showed little effect (Familiar–Agent $0.58$ vs.\ Unfamiliar–Agent $0.56$; MWU $p=0.721$). Overall, agent initial queries were directionally closer to those of more experienced participants, but the differences were not statistically reliable at this sample size.

\begin{figure*}[htb!]
    \centering
    \begin{subfigure}[t]{0.49\linewidth}
        \centering
        \includegraphics[width=\linewidth]{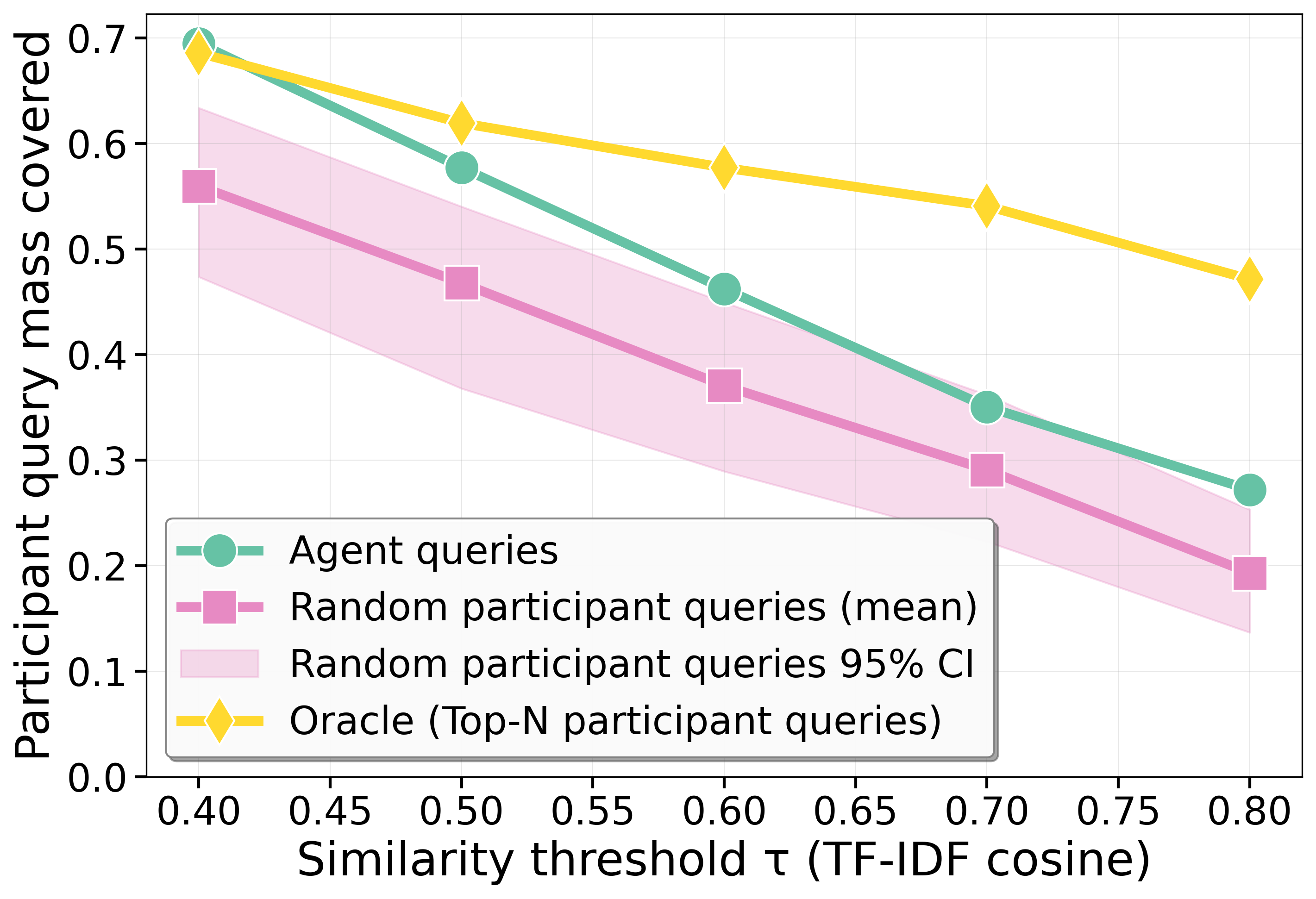}
        \caption{Coverage vs similarity threshold \(\tau\).}

        \label{fig:coverage_A}
    \end{subfigure}\hfill
    \begin{subfigure}[t]{0.49\linewidth}
        \centering
        \includegraphics[width=\linewidth]{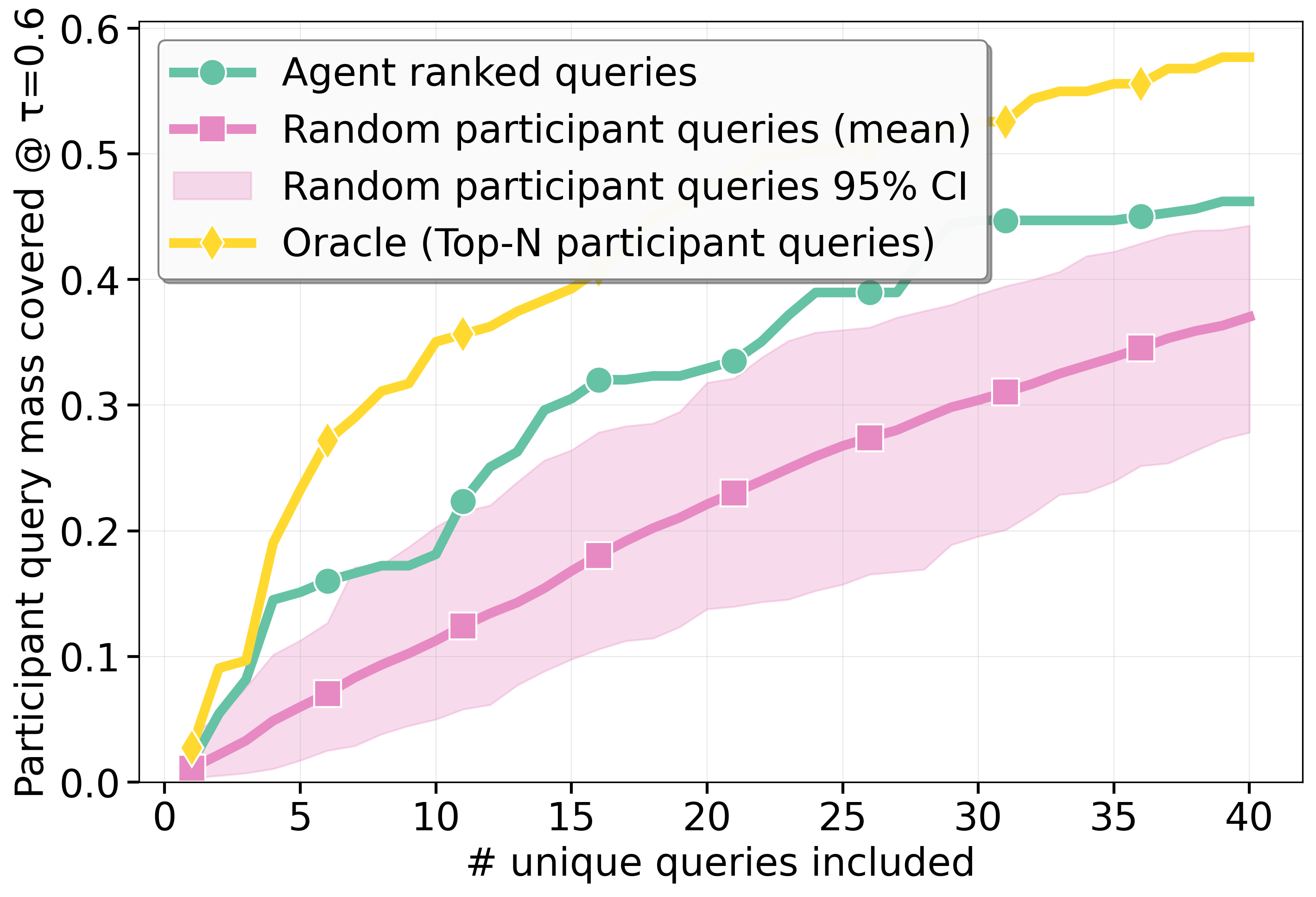}
        \caption{Cumulative coverage at \(\tau=0.6\).}
        \label{fig:coverage_B}

    \end{subfigure}
    \label{fig:coverage}
\caption{Coverage of agent queries within the participants' query space. Curves compare agent (green), a \emph{size-matched random subset of unique participant queries} (pink; mean with 95\% bootstrap CI shaded), and a top-$N$ frequency oracle (gold). \emph{(a)} $N$ fixed to the total number of unique agent queries ($N=40$). \emph{(b)} Each curve adds $k$ queries, with $k$ varying up to $N_{\max}=40$. Both panels use the full query set (TE).}\end{figure*}

\paragraph{Full-query distributional analysis.}
Using all queries from the TE set, we examined how well agent queries covered the participant query space. A majority of agent queries fell within the dense middle of the participant-query distribution: 57\% at or above the median typicality (participant baseline: 50\%). This suggested that a compact agent vocabulary could cover much of the participant query space. Consistent with this, agent queries covered a larger share of the frequency-weighted participant query distribution than a \emph{size-matched random subset of unique participant queries} (Fig.~\ref{fig:coverage_A}), while remaining below the \emph{top-\(N\)} oracle. At \(\tau=0.6\), agent queries covered 46\% of participant query volume, versus 37\% for the random participant-query subset (95\% CI [30\%, 44\%]), while the top-\(N\) oracle reached 58\%. As \(\tau\) tightened, coverage declined for all methods. Fixing \(\tau=0.6\), the efficiency panel (Fig.~\ref{fig:coverage_B}) showed that cumulative coverage rose quickly and remained between the random participant-subset baseline and the top-\(N\) oracle.

Together, these analyses suggested that the queries generated by agent runs were broadly aligned with those produced by study participants.

\begin{figure*}[t!]
    \centering
    \begin{subfigure}[t]{0.4\linewidth}
        \centering
        \includegraphics[width=\linewidth]{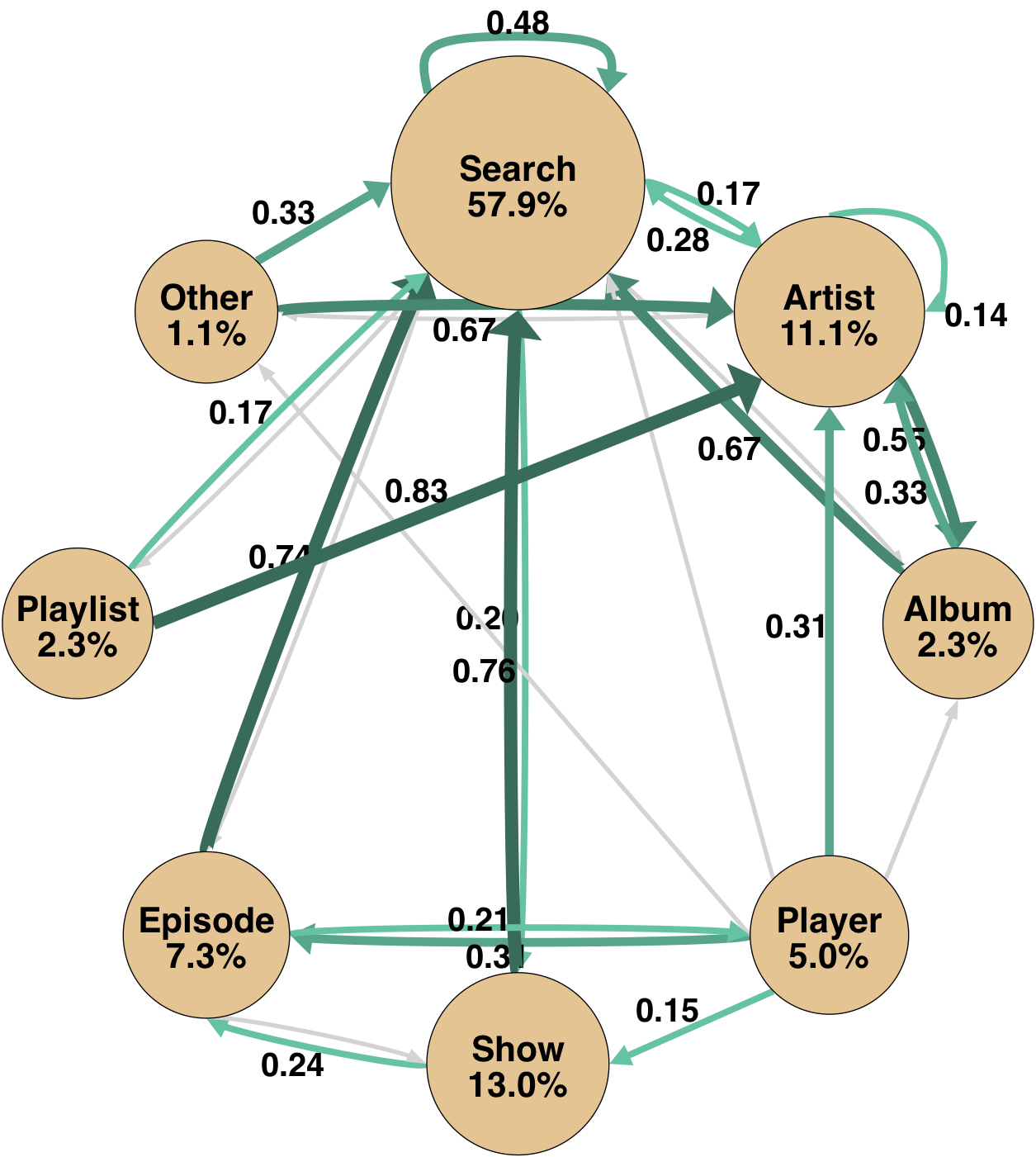}
        \caption{Agent runs}
        \label{fig:graph_agents}
    \end{subfigure}\hfill
    \begin{subfigure}[t]{0.4\linewidth}
        \centering
        \includegraphics[width=\linewidth]{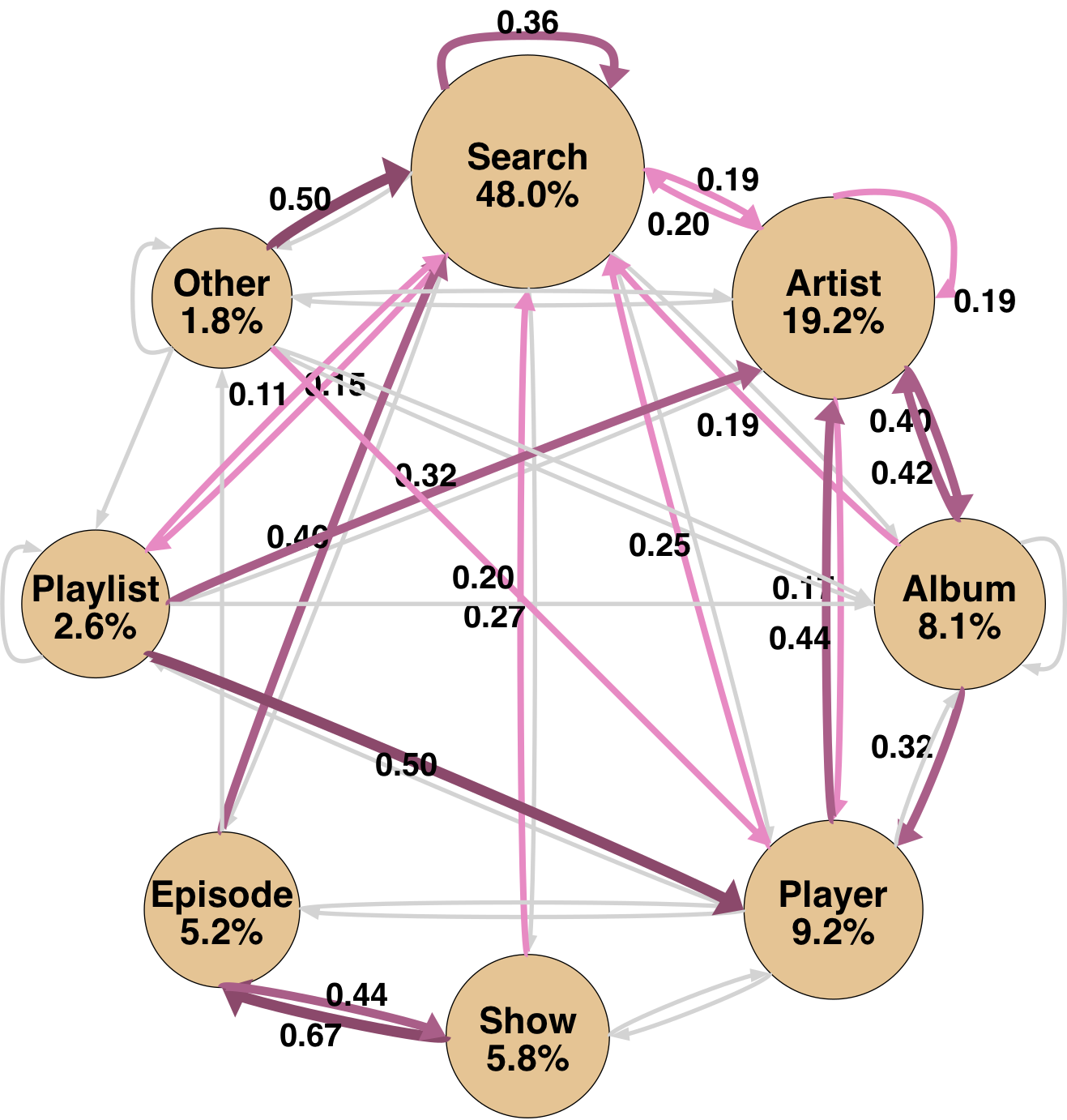}
        \caption{Study participants}
        \label{fig:graph_users}
    \end{subfigure}
  \caption{Macro-averaged transition graphs by cohort (TE set). Node size: visitation share; edge width: transition frequency. For readability, the participant-only \texttt{track} node is omitted from the plots but included in all metrics. Shown are transitions with probability $\ge0.05$; transitions with $<0.1$ probability are {\color{gray}in gray}.}
    \label{fig:graphs}
\end{figure*}

\subsection{Navigation Behavior}

\textit{Search-centric vs.\ content-centric.}
Macro-averaged transition graphs (Fig.~\ref{fig:graphs}) revealed a \textbf{search-centric navigation strategy for agent runs} in contrast to a \textbf{more content-centric, exploratory strategy for participants}. Agent runs showed high search persistence (48\% self-loop probability) and concentrated transitions such as \texttt{playlist}$\rightarrow$\texttt{artist} (83\%) and \texttt{show}$\rightarrow$\texttt{search} (76\%). Participants branched more evenly from the same nodes, for example from \texttt{playlist} to \texttt{player} (50\%), \texttt{artist} (32\%), and \texttt{search} (11\%), and from \texttt{show} to \texttt{episode} (67\%) and \texttt{search} (27\%). Overall, agent runs followed more deterministic, goal-oriented paths, whereas participants were more exploratory.

\paragraph{High-frequency transitions.}
Across tasks, agent runs and study participants shared many of the highest-traffic transitions. Using top-$k$ Jaccard overlap, at $k=10$ the pooled (\emph{micro}, CS) overlap was 0.54 (roughly 7 of the top 10 edges) and the task-equal (\emph{macro}, TE) overlap was 0.44 (roughly 6 of the top 10). As $k$ increased and lower-frequency edges entered, macro overlap declined (to 0.29 at $k=20$), while micro overlap remained higher (0.54 at $k=20$). This difference reflected the aggregation strategy: pooled counts were dominated by frequent hub transitions (e.g., \texttt{search}$\leftrightarrow$\texttt{artist}, \texttt{artist}$\rightarrow$\texttt{album}), whereas task-equal averaging highlighted task-specific branches where agent runs remained more concentrated and participants explored more alternatives.

Overall, the agent and participant traces shared a common navigation backbone, with differences primarily appearing in lower-frequency, task-specific edges. This indicated that while agents and participants often traversed the same high-level interface paths, their detailed navigation behavior diverged.

%% file: tables/effort.tex
\begin{table}[t]
\centering

\setlength{\tabcolsep}{4pt}
\renewcommand{\arraystretch}{1.05}
\begin{threeparttable}
\caption{Task success and effort by cohort (CS Set; all tasks).
Success is reported as completed tasks (s/n), percentage, and 95\% CI.
Time is mean $\pm$ SD (s); actions are median [Q1, Q3].
Statistical comparisons are reported in the text.}
\label{tab:effort-metrics}
\begin{tabularx}{\linewidth}{@{}l c c c c@{}}
\toprule
\textbf{Cohort} & \textbf{Success s/n} &
\textbf{Success \% [95\% CI]} &
\textbf{t (s) } &
\textbf{\#Actions} \\
\midrule
   Agents & $22/39$  & $56.4\ [41.0, 70.7]$ & $1076\pm760$ & $11\ [ 9,19]$ \\
    Participants  & $80/150$ & $53.3\ [45.4, 61.1]$ &  $209\pm175$ & $18\ [12, 30]$ \\
    \midrule
    \multicolumn{5}{c}{Participants' Cohorts}\\
    \midrule
    Expert  & $39/62$ & $62.9\ [50.5, 73.8]$ & $150 \pm 107$ & $18\ [12, 25]$ \\
    Regular & $41/88$ & $46.6\ [36.5, 56.9]$ & $251\pm200$ & $21\ [15, 38]$ \\
    Familiar & $45/75$ & $60.0\ [48.7, 70.3]$ &$184\pm172$ & $19\ [13, 32]$ \\
    Unfamiliar & $35/75$ & $46.7\ [35.8, 57.8]$ & $237\pm177$ & $21\ [14, 38]$ \\
\bottomrule
\end{tabularx}
\end{threeparttable}
\end{table}

%% file: sections/7_practical_lessons.tex
\subsection{Practical Lessons for Validating GUI Agents in Production}

Our results show that outcome alignment does not guarantee behavioral alignment, highlighting the need for behavior-aware evaluation when using GUI agents as proxies in production search systems. For applied settings, we recommend:

\begin{itemize}
\item \textbf{Go beyond success.} Complement task success and effort with trace-level behavioral measures such as query formulation and navigation patterns.
\item \textbf{Validate before proxy use.} If agent traces inform evaluation or optimization, compare agent and human traces on the same tasks to assess behavioral alignment.
\item \textbf{Audit for shortcuts.} Check instruction compliance and shortcut behavior (e.g., prior-knowledge answers) and report exclusion rates.
\end{itemize}

%% file: sections/8_limitations.tex
\subsection{Limitations}

Our study demonstrates the proposed evaluation framework in a single application domain and with one GUI-agent configuration. Although the framework is model-agnostic and applicable to other agents and systems, the specific behavioral patterns observed here may vary across applications, interfaces, and agent architectures. Replicating the study across additional domains and agents would further test the generality of the framework.

As the study was conducted in a live production system, interface elements and content evolve over time, making exact replication of interaction states difficult. This trade-off reflects the goal of evaluating agent behavior under realistic multi-hop information-seeking conditions rather than within a fixed benchmark environment.

%% file: sections/9_conclusion.tex
\section{Conclusions}

We introduced a trace-level evaluation framework for analyzing how GUI agents interact with search-enabled applications relative to human users and demonstrated it in a controlled study within a production audio-streaming application. In this study, 39 participants and a state-of-the-art GUI agent performed multi-hop information-seeking tasks, enabling comparison of interaction traces across task outcomes and effort, query formulation patterns, and navigation behavior.

Our results show that alignment in task outcomes and queries does not necessarily imply alignment in interaction behavior. While the agent achieved success rates comparable to participants and produced similar queries, its navigation strategies differed substantially. Agent runs followed more search-centric, low-branching patterns, whereas participants engaged in more content-centric and exploratory navigation. These findings illustrate how trace-level analysis reveals differences that remain hidden under success-based evaluation. Moreover, the proposed framework provides a practical way to diagnose such differences, supporting behavior-aware evaluation of GUI agents in production search systems.

Directions for future work include integrating human-like exploration strategies into GUI-agent architectures, developing benchmarks that capture richer behavioral metrics beyond task success, and designing training objectives that balance correctness with fidelity to human interaction patterns.